%% file: fnet-article.tex
\documentclass[journal,a4paper]{IEEEtran}
\usepackage[utf8]{inputenc}

\usepackage[english]{babel}

\usepackage[T1]{fontenc}

\usepackage[displaymath,switch]{lineno}

\usepackage{microtype}

\usepackage{url}
\usepackage{siunitx}
\usepackage{graphicx}

\usepackage{amsmath}
\usepackage{mathtools}

\usepackage[capitalise]{cleveref}
\Crefname{figure}{Fig.}{Figs.}
\Crefname{equation}{Equation}{Equation}
\crefname{equation}{}{}

\usepackage{todonotes}

\newcommand{\delaytodo}[2][]{}

\usepackage{booktabs}
\usepackage{multirow}
\usepackage{bigdelim}

\usepackage[mode=image]{standalone}

\usepackage{tikz}
\usetikzlibrary{calc,arrows,arrows.meta,decorations.markings,positioning,shadows,shapes.geometric}%

\usepackage{pgfplots}
\pgfplotsset{width=8cm,compat=1.14}

\usepackage{bytefield}

\usepackage{textgreek}%

\usepackage{bold-extra}

\tikzstyle{busconn} = [thick, decoration={markings,mark=at position
	1 with {\arrow[scale=1, thick]{open triangle 60}}},
	double distance=4pt, shorten >= 5.7pt,
	preaction = {decorate},
	postaction = {draw,line width=4pt,white,shorten >= 5.3pt}]
\tikzstyle{innerWhite} = [semithick, white,line width=1.4pt, shorten >= 4.5pt]

\DeclareSIUnit\bps{bps}%
\DeclareSIUnit\Sa{S}%

\ifdefined\LINENUMBERS

\else

\fi

\newcounter{resumelistcounter}

\begin{document}

\bstctlcite{IEEEexample:BSTcontrol}

\title{Fakernet---small and fast FPGA-based TCP and UDP communication}

\author{
  H.~T.~Johansson, %
  A. Furufors,
  and~P. Klenze%
\thanks{H.~T.~Johansson is
  and A. Furufors
  was with the Department of Physics, Chalmers University of Technology,
    SE-412 96 G{\"o}teborg, Sweden.}%
\thanks{P. Klenze
  is with the Department of Physics, Technical University Munich,
    85748 Garching, Germany.}%
\thanks{The work of H. T. Johansson was supported by
  the Swedish Research Council,
  the Scientific Council for Natural and Engineering Sciences
  under grant 2017-03839 and
  the Council for Research infrastructure
  under grant 822-2014-6644.}%
}

\IEEEpubid{0000--0000/00\$00.00}

\maketitle

\ifdefined\LINENUMBERS
\linenumbers
\fi

\begin{abstract}
  \input{fnet-abstract}
\end{abstract}


\IEEEpeerreviewmaketitle

\section{Introduction}

\IEEEPARstart{I}{n} the at least 100 years \cite{PhysRev.13.272} of
data acquisition history \cite{Meschi2015}, a common theme is the
simplification of data collection.
This work is no exception and is motivated by developments in data
handling in nuclear and particle physics experiments.
However, its applicability is not limited to those fields.
The overall task of a data acquisition is to transport data from
front-end modules to permanent storage and on-line analysis.
Along the way, data from multiple systems is merged, either arranged
by sequence numbers, or sorted by time-stamps, or both.
The later links in the transport chain are typically already realised
as Ethernet networking using commodity hardware, while the front-ends
are custom boards produced in small volumes.
This is also seen in the cost of the various parts as clear
differences between cheap and expensive, with the majority of costs
associated with front-ends and other custom equipment.
An efficient way to reduce overall system cost is to reduce the
amount of custom and low-volume equipment.
Almost all front-end boards have in common that some
field-programmable gate array (FPGA) is present to control the
digitisation stages.

For many front-end systems, manufacturers have designed custom data
buses, that as end-points have receiver adapter cards using the
standard PCIe bus, which thus become the transition point to commodity
hardware.
The present development aims to move this transition point up to the
front-end board themselves, by leveraging the
presence of an FPGA together with using ordinary Ethernet equipment.
The means is a simple and just-enough conforming implementation for
FPGAs of two Internet Protocols (IP) \cite{RFC0791}: Transmission
Control Protocol (TCP) \cite{RFC0793} and the User Datagram Protocol
(UDP) \cite{RFC0768}, such that the FPGA can work as a part of local
Ethernet segments.
Since FPGAs already are available on most front-end boards, the only
additional equipment needed is a physical layer (PHY) chip and
suitable connector, e.g. an 8P8C (RJ45) connector.

\IEEEpubidadjcol

Note that the design of custom PCIe end-point adapters is not easy.
As they appear in later stages of the converging data transport chain
than the front-ends, more data needs to pass each unit, thus needing
faster signalling to fulfil the higher bandwidth requirements.
Custom OS hardware drivers using e.g. direct memory access (DMA)
techniques often also need to be developed and maintained.
Contrast this with the use of well-tested commodity network adapters
and drivers, where the costs of both hard- and software development
are amortized over millions of
users\delaytodo[inline]{\cite{networkadaptersales}}.
\delaytodo[inline]{Ref to sales amount of network adapters.}

To estimate reasonable bandwidth requirements at the front-end,
consider an 8-channel board, connected to a high-rate sensor
which produces as much as 1 MHz of hits per channel.
Note that this is not the sampling speed, but the rate of interesting
signals.
If each hit results in 10 bytes of data, such a front-end would
generate 80 MB/s, i.e. well within the capability of 1 Gbps links.

The basic idea behind Fakernet is to use the fact that commercial
network equipment and general purpose OS network stacks (as found in
GNU/Linux, BSD, Mac, Windows) are designed to work together with with
almost anything, and thus also are able to interoperate with very
simple peers.

\delaytodo[inline]{test the system on the public internet, just to see if it can
handle routing.}

\delaytodo[inline]{test with some BSD network stack.}

\delaytodo[inline]{test from a Mac?  A windows machine?}

The article is structured in the following way:
First, available solutions are reviewed,
followed by the simplifying assumptions the current work is based on.
The operating principles and building blocks of Fakernet are then
discussed.
This is followed by a description of the VHDL module interface and
timing closure considerations, as well as the PC client interfaces.
Finally, the performance of the implementation is benchmarked.

\section{Overview of available solutions}

Custom buses in data acquisition systems are generally designed as
daisy chains in order to limit the number of connectors at the
receiving PC.
This is illustrated in the lower part of \cref{fig:topology} and seen
in \cref{tab:custom_buses}.
The listed buses have been chosen as they have been developed for some
level of generic use, and not only for single systems.
In Ethernet-based topologies the daisy chain is avoided by the use of
switches.

\input{figure_topology}

The use of TCP/IP directly at high-performance front-end boards for
data acquisition purposes require high transfer speeds, low FPGA
resource consumption and simple interfaces.
Our experience in developing and handling such highly customised
systems as data acquisitions are, is also that open and free
availability of source code benefits both the design phase and
long-term maintenance.

Commercial FPGA manufacturers offer TCP/IP cores for their respective
platforms, but by being generic implementations, they by necessity
require the user circuit to handle more of connection management
leading to more complicated interfaces.

\Cref{tab:fpga_tcp_ip_impl} list TCP/IP implementations for
FPGAs found in the literature.
Most generic implementations use more resources than systems dedicated
to only transmit TCP data, with the exception of \cite{Uchida}.
\cite{Uchida} also has a dedicated control interface, while that would
have to be constructed on top of the generic packet handling for the
other implementations.

\delaytodo[inline]{why your readout should not be real-time;
  hot take: synchronous readout systems scale badly and are inherently
  complex because of all the real-time requirements}

\input{table_custom_buses}

\input{table_fpga_tcp_ip_impl}

\section{Simplifying restrictions}
\label{sec:simplifying}

The overall goal is to keep the circuit small.
Two requirements guide the design:
\begin{enumerate}

\item %

  Except for the front-end boards, all other components in the data
  transfer chain shall be standard commercial, off-the-shelf (COTS)
  equipment.
  In particular, this includes cables, switches and network
  adapters.

\item No special OS/kernel drivers shall be needed for the receiving
  computer.

  Communication shall be possible using normal user-space tools, in
  particular using the BSD/POSIX \texttt{socket} interface for TCP and
  UDP.
  Administrator privileges shall only be needed for one operation on
  the receiving computer: to configure the IP address of the internal
  network interface connecting to Fakernet.

\end{enumerate}

Speed is only a concern for data transport out of
the front-end board.
There it is desirable to achieve line-speed, at least
up to 1 Gbps.

While the above requirements lead to an implementation which adheres
to the basic principle operation of Ethernet, IP, TCP and UDP, they
still allow a number of simplifying restrictions and assumptions.
This is possible since in data acquisition scenarios the entire
hardware chain is under control, and thus the implementation only need
to work in some well-defined configurations, not any.
\begin{enumerate}

\item Data is only transmitted over TCP, never received.

\item All accepted and transmitted packets have an even number of
  octets.
  This restriction is no issue for UDP data, since the sending
  application chooses the payload size.
  It could be a problem for receiving TCP data (which is not done);
  but likely to not be a problem in practice.

  \setcounter{resumelistcounter}{\value{enumi}}
\end{enumerate}
Since operation only is intended on non-routed local (private)
segments, the following restrictions also have limited impact:
\begin{enumerate}
  \setcounter{enumi}{\value{resumelistcounter}}
  
\item Only IPv4 is supported.   %

\item No IP, TCP or UDP options are supported.
  In particular, TCP window scaling is not used, only ignored.

\item \label{nobottleneck} The network should have no bottlenecks
  between Fakernet and the receiving PC.
  All bandwidths at and after successive switch stages should be
  larger or equal to the sums of the connected, and worst-case used,
  total incoming bandwidth from front-end boards.
  This would e.g.\ mean to use 10 Gbps links after a switch
  with 1 Gbps front-end connections.

\item Simpler means than normally employed to govern TCP bandwidth
  control and retransmission can be used thanks to low delay of
  small/short-distance local network segments.

\item No security measures are implemented, e.g.\ against spoofed packets.

\end{enumerate}

Assumptions 5 and 6 do not impact correct operation, only performance
is affected when they are violated.

\section{Operating principle}

In the following, we call the other network endpoint that interacts
with Fakernet for the PC.
The other code in the FPGA, with which the Fakernet firmware
interacts, is referred to as the user (circuit) code.

The basic operating idea is that Fakernet never generates any packets
from scratch, or on its own initiative, except for TCP data.
Instead, all emitted packets are constructed on-the-fly by modifying
incoming packets.
This is possible and straight-forward since all response packets
(except TCP data transmissions) have the same length and general
layout as the packets they respond to.
Fields are therefore either copied or modified directly, or with
swapped locations for source / destination items, such as media access
control (MAC) and IP addresses and port numbers.
The structures of the handled packet types are shown in
\cref{fig:matryoshka}.

These packet transformations are performed directly in the input
packet parser, which writes the response while the input words of a
packet are inspected.
The changes are easily performed, as the outgoing responses have
the same length as the incoming packets.
For each incoming two-octet word, a two-octet word is written to the
response memory, but possibly at a different location (which is
handled as an offset), see \cref{fig:word_copy}.
Since it at the beginning of an incoming packet is not known what kind
of packet it is, all available destination memories are written (c.f.\
\cref{fig:overview}).
If the incoming packet pass all checks and the generated response
therefore shall be transmitted, the created packet in the relevant
memory is marked for transmission.
This applies directly to Address Resolution Protocol (ARP) \cite{RFC0826}
and Internet Control Message Protocol (ICMP) \cite{RFC0792} responses.
A response UDP packet is further modified with actual data responses
and success markers when going through the unit performing the
register access before the response is fully ready for transmission.
TCP packets are generated in a separate unit, with or without data,
but based on a template packet taken from the initial SYN packet for
all header fields.
Thus the TCP generator can be simple and only need to modify the
length field.

The input to the circuit is two-octet words (16 bits) with a flag
telling in which clock cycles they represent new data.
The output is also provided as two-octet data words, and the outside
link-layer circuit must flag every cycle when such a word was consumed
by the transmission.
This allows the circuit to operate at any frequency that is fast
enough to handle the incoming words.
Any clock-domain crossing is the responsibility of the user.

The concept of dropped packets is used extensively.
If a resource is temporarily unavailable, e.g.\ the memory for storing
a response packet is occupied, then the corresponding incoming packet
will simply generate no response, i.e.\ effectively be ignored.
Consequently, the internal state is not updated in these cases.
Thus, this central feature of the IP stack which might
sometimes be perceived as a draw-back is instead used as an advantage,
considerably simplifying the circuit design.
Note that for TCP, dropped packets indicate bandwidth limitations.
Fakernet never drops well-formed TCP packets.

\input{figure_matryoshka}

\subsection{Clock frequency}

The Fakernet circuit is designed to handle 16 bits (two octets) of data
each clock cycle.
Compared to handling one octet per cycle, this reduces the timing
requirements considerably as only half the clock frequency is needed
compared to handling one octet per cycle.
Going further and processing input words in 32-bit chunks would
however require word-shuffling logic, since some reordering between
the incoming and outgoing words would no longer be simple, due to
mixing within 32-bit aligned boundaries.
Also, the Ethernet frame header has an odd number of 16-bit entries,
which would cause misalignment.

\input{figure_word_copy}

\subsection{Hardware address and IP address}

In order to operate together with COTS equipment, each Fakernet
system must have a unique IP and hardware (MAC) address.
Providing this is outside the scope of Fakernet.
Is is given by the user circuit (through the VHDL interface), and
unless originating from an unique-ID chip, it is suggested to use a
hardware address marked as locally-administered (by setting the
second-least significant bit of the first octet, the U/L bit)
\cite{EUIguidelines}.
A simple identification number can be used for the remaining parts of
the MAC address, since it is only seen in the local private network.
When multiple front-end cards are connected to the same PC, this
identification need to be unique, and thus often require some kind of
switch or other selector on the front-end hardware.

The IP address can follow the same scheme: use the identification
number for the least significant bits or octets, and select a private
range for the most significant octets.
We suggest 172.x.y.z, since it seems more common that 192.168.x.y or
10.x.y.z are used for local lab networks, thus becoming the outside
network for the PC in data acquisition environments.

Packets which are not for the chosen MAC or IP address are ignored.
Any destination MAC address is however accepted for ARP requests.

Note that using DHCP to assign an IP address would not provide any
simplification:
A unique MAC address would still have to be provided.
Moreover, the system would need to generate DHCP queries, and
interpret the responses.

\input{figure_overview}

\section{Building blocks}

Operation is governed by a few finite state machines (FSMs).
In-between them, packet data is passed uni-directionally using
dual-ported RAM blocks, each with one writer and one reader.
Along with each memory, a control block holds information whether the
memory has packet data for the consuming end, and how much.
An overview of the entire Fakernet circuit is shown in
\cref{fig:overview}.

A large fraction of the functionality is in the input-packet FSM.  It
must be able to deal with almost any kind of input data.
It takes the easy approach of only advancing the parser state for
input words that are acceptable at the current location.
Any unacceptable word will return the FSM to the idle state, where it
waits for the start of the next packet, see \cref{fig:in_fsm}.
Incomplete response packets are simply not marked for consumption, and
thus never sent.

As the other state machines deal with data or packets that have
already been validated, data-dependent failure is not possible and
they generally go through all motions after they have started the
processing of a packet.

\input{figure_in_fsm}

\subsection{ARP and ICMP handling, MAC and IP address lookup}

For the system to interact with normal IP stacks, it must respond to
ARP requests, in order to provide its MAC address when asked who has
its IP address.

Furnishing the response is simple: as for all other packets, for the
destination MAC and IP address, replace with the incoming source
address.
Fill out the local Fakernet values as the source for MAC and IP
address, and change the ARP operation code in the packet to
a response operation code.

While it is not strictly necessary for the system to respond to ping
packets (ICMP type 8; echo request), doing so simplifies operations
debugging considerably.
Since the UDP and TCP handlers are much more restrictive and only
respond to packets with certain payloads, ICMP echo request/reply
makes it possible to check the MAC and IP address assignment with the
ubiquitous user-space tool \texttt{ping}.
Furthermore, as the handling of IP headers is also needed for UDP and
TCP packets, the additional overhead to handle ICMP echo requests is
very small.

The same intermediate memory is used for ARP and ICMP response packets
(as well as some initial UDP sequence reset responses).
This is not a problem, as it only
means that if several such packets arrive in rapid succession, before
the first response has been sent, they are dropped.

\subsubsection{RARP (unimplemented)}

In case the front-end board has a unique-ID chip to provide the
hardware address (MAC), but no way to directly select a unique IP
address, RARP could be used to provide the IP address.
This would be needed to avoid potential collisions, since the IPv4
address cannot be directly based on the ID chip.
This is due to that fewer bits are available in an IPv4 address than a
MAC address.  ID chips are generally designed to fully provide the
latter.

While the PC does most of the work by running a RARP daemon, one way
for the Fakernet circuit to do its part would be to have a prepared
RARP request packet in a memory, where only the own hardware address
need to be replaced upon transmission.
Alternatively, while the IP address is still unconfigured, any
incoming ARP request, which otherwise would be dropped, could be used
as a template to drive the RARP request generation in the input FSM.
Thereby the usual ARP response is repurposed to become a RARP request,
since the formats are very similar.

\subsection{Output generator}

While idle, the output FSM looks for packets that are ready to be sent
in any of its source memories.
TCP packets are sent with lowest priority, in order to not starve
control requests via UDP or other low-volume ARP/ICMP replies.

The output generator first transmits the Ethernet preamble (7 octets
of 0x55 and one 0x5d octet), followed by the packet data.  It ends
with the frame check sequence (FCS) checksum and an inter-packet gap
of 12 octets before selecting any next packet to transmit.

\subsection{Checksums}

Checksums are a necessary ingredient to make the system operate
together with standard hardware and IP stacks.
They are also one of the great strengths of Ethernet and the IP-based
protocols.
All checksums are verified for all incoming packets.
Any failing packet is simply ignored.

The checksums are also a reason to implement the MAC layer directly in
the FPGA.
With the Ethernet checksum verified and produced directly in the FPGA,
all transmissions on the front-end board are also protected against
corruption.

The in-packet checksums (IP header, TCP header and UDP) are calculated
when the packets are prepared, and written to memory shortly after the
relevant section is complete.
These are all 16-bit, ones' complement, wrapped-accumulator sums of the
data words, thus their evaluation fit well with the handling of two
octets each (active) clock cycle.

The Ethernet FCS is generated on-the-fly by the output FSM, and is
available for transmission immediately after the last double-octet of
the packet payload or padding has been sent.

(Not-yet-implemented) In case a parity error is detected while reading
an internal memory during packet transmission, the last octet of the
checksum is inverted.
This will cause any receiving unit (switch or network adapter) to drop
the packet, since the transmitted FCS does not match its actually
transmitted wrong data.
Since the packet when the error is detected already has been partially
transmitted, this is a clean way to propagate the fault condition.

\subsection{UDP handling---register access}

The Fakernet circuit provides an address+data interface to perform
control operations of other user circuits in the FPGA.
Access is provided over UDP, but implemented so as to
give reliable operation.
This is done by means of a sequence number, which must be incremented
by one for each new access from the PC side.
Unacknowledged UDP requests must be retransmitted after some
timeout, until a response is received.
If the previous sequence number is seen, the system retransmits the
previous response (without performing the actual access again).
This handles both the cases of dropped request and response packets.
When retransmitting, the PC side must actually send the original
request unmodified, since if the original request packet was lost, the
actual access has not happened, and will be performed due to the
arrival of the retransmitted packet.

\input{figure_udp_reg_acc}

In order to allow the PC to have a few independent control sequences
performed simultaneously, several (by default two)
channels are implemented, each responding on a separate UDP port.
To activate such a channel, the PC must first pre-request (arm) a
reset of the sequence number, and then reset it, after which it can
perform operations.
The overall exchange of packets is shown in \cref{fig:udp_reg_acc}.
The arm-reset sequence is two-step to avoid resets due to single
spurious packets.
Channels which have been recently used cannot be reset at all, within
a user-configurable timeout (default about a second), thereby
somewhat protecting the current channel user.
For each active channel, the system keeps track of the IP address and
UDP port of the client PC side, thus preventing other processes from
interfering with an access channel by mistake.

These mechanisms provide no protection against register access packets
from different channels to be performed between each other.
But each packet is handled uninterrupted.

\subsubsection{Idempotent actions}

Access channels are a scarce resource.
For non-critical and idempotent actions, like
reading status or counter registers, it is not necessary to acquire
and hold a dedicated channel occupied.
Therefore, the first UDP port ignores the sequence number and also
does not check or retain the PC side IP address and port number.
It thus can handle and perform all requests (when the response memory
is not occupied), regardless of the number of clients.

\subsubsection{Internal registers}

The register interface also give access to some Fakernet-internal
registers and (optional) debug counters.
This interface is selected with the highest (28th) address bit.
The counters have a light-weight (resource conserving) implementation
with the actual values stored in a RAM block.
Single-bit flags are set for each event that shall be counted, and an
updating process adds one of the set flags to its counter every fourth
clock cycle.
This is accurate since the counted events occur even more seldomly than
the cycling of the update process.

Some important status bits are provided directly at the start of each
UDP packet response, and are thus always accessible without acquiring an
access channel.
These mark which UDP access channels are in use, and
the current TCP connection state.
It also reports any faults due to attempts at overfilling the TCP data
buffer, or detected parity errors.

\subsection{TCP handling}

A TCP connection is initiated from the PC side, by the transmission of
a SYN packet.
If the TCP state is unconnected, the received packet
will be used as a template for the TCP output preparation state
machine, which will first be asked to generate a SYN-ACK packet (with
no data).
If, however, the TCP state is connected, the connection attempt is
ignored.
Upon reception of the following ACK packet from the PC side, the TCP
connection has been established and transmission of data commences.
Before each TCP connection attempt, the Fakernet TCP state must be
reset with a dedicated internal register access via UDP.
In particular, only one SYN-ACK response packet is sent per reset.
TCP options in the incoming packets are ignored.
No options are sent.
For simplicity, almost no bandwidth control is employed (c.f.\
\cref{sec:bwretransmit}); data packets are generated as soon as a TCP
output memory is free.

Note that establishing a new Fakernet TCP connection requires a reset,
which clears the data buffer.
Thus data will almost certainly be lost when a connection is lost.

The data transmission is handled by three actors:

\begin{enumerate}

\item The TCP state and control, which keeps track of the current send
  base (how far the PC side has acknowledged data), and the
  current send front (how much data has been sent at least once).
  It also keeps track of how much data is available, i.e.\ has been
  filled by the user circuit, and the current acceptance window
  announced by the PC side.
  The data is kept in a circular memory internal to the Fakernet
  system, see \cref{fig:data_memory}.

  The state is updated from either the input packet handling or the
  output packet generation, on completed packets of either kind.
  The state updates only affect each other through the differences of
  the values.
  Whether further transmissions are possible at any given time is
  determined by the differences between the send base and front, and
  available data, as well as the current window.

  The TCP control is also responsible for measuring RTT times
  \cite{RFC6298}, and requesting retransmissions, see \cref{rttmeasure}.

\item Input packet FSM.
  The only validation performed by the input FSM is that the given
  acknowledgement location is within the currently transmitted window,
  i.e. above or equal to the current send base and less or equal to
  the current send front.
  Such acknowledgements are reported to the TCP state.
  No input payload data is handled.

\item TCP packet preparation FSM.
  The headers of the generated TCP packets are based on the template
  packet recorded from the first (SYN) packet of the connection.
  Packet data is generated into one of two temporary memories in a
  ping-pong fashion, which are consumed by the output packet sender
  FSM.

  Each packet must be prepared into a memory, since the TCP checksum
  depends on the payload data, but appears before the data in the
  packet, and therefore a two-pass operation is required for actual
  transmission.
  One pass is thus in the preparation FSM, and the other pass in the
  output packet sender FSM.

  Two memories are used such that the implementation can prepare
  another packet while the previous is transmitted, allowing it to
  saturate the network interface by continuous transmissions.

\end{enumerate}

\input{figure_data_memory}

\section{VHDL module interface} %

The interfaces to the user circuit are designed to be simple, with few
signals and little sequence and state handling.

\subsection{Control register interface}

The register access interface gives the address and the values of data
to write or take the read return, as well as flags to indicate the
direction of the operation.
The user code shall respond with a done flag within about 10 clock
cycles.
This should give ample time for the client circuit to perform whatever
pipelined multiplexing or fan-in/out that is required for the access,
without having to construct code which becomes critical for timing
closure.
If no response is given within the allocated number of cycles, the
access is considered as failed and not marked as performed in the
output response packet.

Interface signals:
\begin{enumerate}

\item \texttt{reg\_addr} A 25-bit address for the item to be accessed.

  The four high bits of the 32-bit address word in the access packet
  are used to indicate the direction of access (read/write), and (in
  the response) if there was an actual response, i.e.\ a successful
  access.
  The 28th address bit marks Fakernet internal register access.
  Two bits are reserved for future use.
  
\item \texttt{reg\_data\_wr} Data value for write access.

\item \texttt{reg\_data\_rd} Return data value for read access.

\item \texttt{reg\_write} Flag indicating a write operation.

\item \texttt{reg\_read} Flag indicating a read operation.

\item \texttt{reg\_done} Return flag indicating a successful operation.

\end{enumerate}

The read and write flags are only active for one clock cycle.
If several cycles are needed, the user must pipeline the flags.
The address and write values are held, so need not be latched.

\subsection{Data transmission}

Data to be transmitted via TCP in a streaming fashion is given from
the user circuit to Fakernet in groups of arbitrary length.

The data input interface allows the producing entity to generate data
words in any order within a commit group.
It can e.g. first write payload data of yet unknown length, and as a
final step write a header.
After the group is completed, it is committed for transmission in one
go.
The same data word can be overwritten many times, but committed data
can not be modified.
It is the responsibility of the user to write each data word.
Failure to do so causes old data to be transmitted.
A word can be written to a group in the same cycle as it is committed.
Note that commit group boundaries are not retained or marked in the
streaming data.

The circular data buffer is handled such that
the maximum offset the user can give for writing never can
write data words in the part of memory that is currently being
transmitted (i.e. not yet acknowledged).
A flag indicating that free space is available is provided to the user
circuit.
In case the user nevertheless tries to commit (or write) despite
absence of free space, it will be flagged as an buffer overflow, and
no further writes will be handled.
\delaytodo[inline]{We can allow writes, but not commits.  Easier user code.
  Keep as is, i.e. writes are also not allowed, but in actual code and
  readme allow for the other case.}
Such behaviour is a violation of protocol and thus indicates a bug in
the data producer.
This can only be reset by restarting the TCP session.

For simplicity, free space is no longer reported when less than three
times the maximum write offset is available, but an access is only
considered as overflow when a write would be beyond one times the
offset.
This allows the data-producing user circuit to start a new group and
complete it, if the free marker was present when the group was
started.
It also allows for the few clock cycles of delay before the user
circuit notices the absence of the free marker, after a commit which
passes the first (three times maximum offset) threshold.
The delay is a few clock cycles, since input and output signals of the
interface are latched twice, so as to not tightly couple the user
circuit and the internal system circuit, which might otherwise lead to
potential timing closure problems at the interface between the
systems.

Interface signals:
\begin{enumerate}

\item \texttt{data\_word}  Data word to be written (32 bits).

\item \texttt{data\_offset} Offset in the current group to write
  the data at.

\item \texttt{data\_write}  Flag to mark that a data word is to be written.

\item \texttt{data\_commit\_len}  Amount of data to commit.

\item \texttt{data\_commit}  Flag to mark that data should be committed.

\item \texttt{data\_free} Return flag marking that space is available
  for at least one more commit group.

\item \texttt{tcp\_reset} Flag marking that the TCP state has been reset.
  The data buffer is also cleared on reset.

\end{enumerate}

\delaytodo[inline]{a reset return flag, to stretch the reset as far as
  needed.  Better make a generic map parameter that tells how many
  cycles to hold a reset.}

\subsection{Ethernet}

The direct interface to the Ethernet hardware (PHY chip) is not part
of Fakernet, but must be provided as some user circuit.
It is also responsible for the input preamble handling,
i.e.\ detecting the start of incoming packets.
Nevertheless, a few examples are provided alongside the source
distribution.
E.g. for an MII interface, this handles the conversion between the
4-bit data nibbles from and to the PHY and the internal 16-bit
interface.

Fakernet input interface:
\begin{enumerate}

\item \texttt{in\_word}  Input two-octet (16-bit) value.

\item \texttt{in\_gotword}  Flag indicating the presence of a new input word.

\item \texttt{in\_newpacket} Flag indicating the beginning of a new
  packet.  It shall be given directly before the first word, i.e.\ at
  the Ethernet start of frame delimiter (SFD).  (Note that the
  preamble and SFD themselves need not be delivered.)

\end{enumerate}
Output interface:
\begin{enumerate}

\item \texttt{out\_word}  Output two-octet (16-bit) value.

\item \texttt{out\_ena} Flag marking that the current word is part of
  the packet or preamble, and not just inter-packet gap filler.

\item \texttt{out\_payload} Flag marking that the current word is part of
  the packet, but not the preamble.

\item \texttt{out\_taken} Return flag from the user PHY interface
  circuit indicating that the output word has been consumed.

\end{enumerate}

Configuration and utility interface:
\begin{enumerate}

\item \texttt{cfg\_macaddr}  The MAC address to be used.

\item \texttt{out\_ipaddr} The IP address to be used.

\item \texttt{slow\_clock\_tick} Pulse to drive the internal RTT
  counters.  Should have a period of \SIrange{0.5}{5}{\micro\second}.

\item \texttt{timeout\_tick} After two timeout ticks, UDP connections
  can be reset.  Suggested to be on the order of a second.

\end{enumerate}

\subsection{Timing closure}

\input{table_resource_use}

\Cref{table:resource_use} shows the Fakernet resource consumption,
with varying number of optional items.
The code is able to run at well above \SI{100}{MHz} on many FPGAs.
This is more than enough to handle a 1 Gbps link.
With octets transmitted at \SI{125}{MHz}, only a \SI{67.5}{MHz}
operating frequency would be required due to the 16-bit interface.
All interfaces to the user circuit are also proactively pipelined,
such that the chance of timing-critical paths appearing during
synthesis at the interface between the module and outside entities is
reduced.
Furthermore, all input and output interfaces can easily be further
pipelined by the user, since there are no single-cycle reaction
requirements between input and output signals, with the exception of
\texttt{out\_taken}.

Note that at an operating frequency of \SI{156.25}{MHz}, it would be
possible to directly use \SI{2.5}{Gbps} Ethernet links.
Support for the TCP window scaling option need not be implemented, if
the larger bandwidth-delay product stay within the \SI{64}{kiB} bound.

As a future development possibility, if the TCP packet preparation and
output state machines are extended to handle data in 64-bit chunks, it
would at the same operating frequency be possible to deliver TCP
output data at \SI{10}{Gbps}.
Together with a rate-lowering buffer stage before the input FSM, no
other logic would need extension, except for TCP window scaling.

\section{PC Client Interfaces}

\subsection{Client TCP interface}

Interfacing with the TCP stream is on a POSIX system done through the
normal BSD \texttt{socket} system calls:

\begin{enumerate}

\item First, the TCP session has to be restarted through a register
  access via the UDP interface, see below.

\item \texttt{fd = socket()} to create a socket file handle.

\item \texttt{connect(fd, ...)} to connect to the IP address and
  designated port.

\item Repeatedly call \texttt{read(fd, buf, size)} to read data.
  
\end{enumerate}

\subsection{Client UDP interface}

The UDP interface likewise uses \texttt{socket} and the usual
\texttt{sendto()} and \texttt{recvfrom()} interfaces on the PC side,
to send and receive individual packets.

A small library with utility functions to establish channel access and
perform the register access is also provided.
It handles the necessary retransmissions after timeout in case of
missing responses.

\delaytodo[inline]{Describe the functions of the library interface.}

\input{figure_udp_flood}

\section{Performance}

The circuit has been tested on a Digilent Arty A7 35T board
\cite{web:artya7}, which provides a 100 Mbps PHY directly connected to a
Xilinx Artix-35T FPGA.
Several performance tests have been conducted to verify the
throughput capabilities.

A built-in data generator controlled by the register access interface
was used to feed the tests.

The internal data generator is a part of Fakernet (but can optionally
be disabled during synthesis to save resources).
When enabled, it can be used in any system to verify (stress-test) the
capabilities of the network hardware.
Furthermore, the normally not modifiable TCP receive window (given be
the receiver) was artificially reduced under control of an internal
register, as well as the maximum length of transmit packets.

The tests have been performed with a Xeon E3-1260L CPU running
single-threaded client code at 3.3 GHz.

\delaytodo[inline]{Xeon E3-1285v6 CPU running
single-threaded client code at 4.5 GHz.}

\delaytodo[inline]{10 Gbps network adapter to reduce PC-side latency?}

\subsection{UDP interface, maximum throughput}

The UDP access protocol is able to perform \SI{450}{kword/s}
(= \SI{1.8}{MB/s}) register accesses with the tested \SI{100}{Mbps}
interface, see \cref{fig:udp_flood}.
Assuming a similar PC-side overhead at \SI{1}{Gbps}, these values
would be \SI{1.2}{Mword/s} (= \SI{4.8}{MB/s}), i.e.\ less than a
factor 3 higher, due to the rather significant CPU overhead.

\subsection{Throughput vs. maximum packet payload length}

\input{figure_maxpayload}

Normally, when data is available, Fakernet transmits using packets
with payload lengths up to 1440 bytes.
\Cref{fig:maxpayload} shows the throughput as a function of an
artificially constrained maximum payload length per TCP packet.
It is seen that Fakernet is able to saturate the network link with
streaming data, in the limit of large, unconstrained packet payloads.

\subsection{Throughput vs. maximum receive window}

\Cref{fig:maxwindow} shows the throughput as function of an
artificially constrained maximum receive window.  This is varied in
order to illustrate the relationship between the link bandwidth $f_\mathrm{BW}$, the rount-trip time
$t_\mathrm{RTT}$, and the amount of not yet acknowledged, in-flight
data $w_\mathrm{in\textnormal{-}flight}$ allowed by the receive window,
\begin{equation}
  \label{eq:se}
  w_\mathrm{in\textnormal{-}flight} = f_\mathrm{BW} \cdot t_\mathrm{RTT}.
\end{equation}

As long as the receive window is large enough to not limit the amount
of outstanding data due to the round-trip time, the network link is
saturated.

Since \SI{120}{\micro\second} of the RTT can be attributed to
transmitting a full-sized packet at the \SI{100}{Mbps} link speed and
thus the RTT should be reduced from \SI{180}{\micro\second} to
\SI{70}{\micro\second} at \SI{1}{Gbps}, it can be estimated that a
window size of \SI{8.8}{kiB} should be enough at this higher speed.
As this is smaller than \SI{64}{kiB}, window scaling is not
needed.
\delaytodo[inline]{The RTT measurement is for single-packet transfers.  Is this applicable?  Obviously --- it matches.}

\input{figure_maxwindow}

\section{Conclusion}

A circuit to enable FPGA-equipped systems to act as direct data
sources using the ubiquitous TCP/IP protocol and cheaply
available commercial hardware has been presented.
The circuit can be directly integrated in front-end electronics, and
only need a PHY interface and suitable connector, e.g.\ 8P8C (RJ45).
It also provides a reliable address+data control interface using UDP
communication.
In neither case, no special drivers are needed for the controlling
computer---all communication use ordinary user-space functionality.

The overall architecture as a data-flow circuit was described, where
the majority of the work is handled already in the state machine
parsing the incoming Ethernet packets.
While the circuit is not a generic TCP or UDP implementation, for the
dedicated tasks at hand---continuous data stream transmission and
control operations---it provides straight-forward and
easy-to-use interfaces to other code in the FPGA.

Performance measurements show that the circuit easily saturates
\SI{100}{Mbps} network links.
The performance behaviour is also well-described when the TCP payload
or window size are artifically reduced.
It was also argued that the circuit would be able to saturate
\SI{1}{Gbps} links, although that has not yet been tested in practice.

Computer code for the FPGA implementation in VHDL and for the CPU
client routines in C are available for download~\cite{web:fakernet} as
open source software.

\appendix

\subsection{Bandwidth and Retransmission}
\label{sec:bwretransmit}

Two important concepts for TCP are flow control, which is about not
overwhelming the other end of a connection, and congestion control to
avoid oversaturating the network in-between.

The Fakernet implementation makes no attempt at explicit congestion
control, which would require it to determine the effective available
bandwidth through the network to the destination.
One of simplifying assumptions in \cref{sec:simplifying} is that the
network has no bottlenecks between the front-end and the destination
network adapter, and thus that no packets are dropped due to
congestion.

Flow control is however handled---it is also more direct.  If the
receiving PC is not able to deal with the data rate, this will lead to
it announcing a smaller or even zero receive window, which is honoured by
the implementation.
After a zero receive window, the sender is responsible to probe the
connection for a non-zero window to eventually continue the
transmission.
This is handled by the RTT timer, which will perform retransmission
even of zero-length data, when a zero receive window has been
announced.

Regardless of the simplifying assumptions, the implementation must
gracefully handle situations where packets are lost, e.g.\ due
to insufficient bandwidth.
This is as usual done by retransmission, where the TCP specification
leaves the details of handling retransmissions to the respective
implementation.
What must be respected is that it is the responsibility of the
transmitter to make further delivery attempts, while \emph{when} to do
it is at its own discretion.

If Fakernet receives a double-same ACK (i.e. three TCP ACK packets
from the receiving end that repeatedly specify the same location),
then one new repeated TCP packet at the current acknowledged point is
generated.
This is typical handling, since multiple ACKs at the same location
indicate that an intermediate (the next) packet in the data stream has
been lost on the way to the receiver; the non-moving ACKs come from
the receiver in response to later data packets.
In addition, if no ACK has been received during twice the estimated
round-trip time (RTT), then also one retransmit packet at the current
ACK point is generated.

Note that this conservative use of retransmissions will cause recovery
of lengthy losses to be rather slow, since generally two RTT will pass
between each retransmit packet.
This is however not the common case in the designed-for topologies,
with over-provisioned bandwidth.
When the cause is an occasionally lost packet, then only that is
missing at the receiver, and a large move in the acknowledge point
will occur after the other end receives the retransmitted packet.
Thus there is also no need to handle selective ACKs, which is a
complicated TCP option to parse and keep track of.
If the cause of packet loss is insufficient bandwidth, the implemented
scheme means that this sender will effectively back off from sending
full speed for a while (until the acknowledge point catches up with
the full-speed sending front).
Thus other senders will have a larger chance getting their data
through the bottleneck.
This effectively constitutes an implicit congestion control.

\subsection{RTT measurement}
\label{rttmeasure}

\input{figure_trickle}

The RTT is measured continuously.
When no measurement is ongoing and a packet is generated at the
sending front (i.e. not for retransmissions), a timing counter is
started and the front location is recorded.
When data is acknowledged beyond the recorded location, a round-trip
has happened, and the counter value is fed to a filter.
If a retransmission is done while a measurement is ongoing, the
measurement is cancelled, since it will be unknown if the original or
the new packet caused an eventual acknowledgement response.
The filter is based on 16 measurements, with the result as the minimum
value of groups of four, where in each the maximum value has been
kept.
The result is used until a new fully filtered measurement is obtained.

For each retransmission due to RTT counter timeout, the RTT estimate
value is incremented by one unit, up to its maximum value.
This is a slower reduction than mandated by \cite{RFC6298}, but is
required since the normal increment of a factor two of the RTT
estimate each retransmission would not work well together with the
retransmissions being the only way in this implementation to recover
from lengthy losses of data.
It still achieves a slow decrease of the retransmission rate.
Thus, if the other end looses track of the connection, bandwidth
consumption will reduce to a trickle, without having to consider
timeouts for when to drop a connection, see \cref{fig:trickle}.
With RTT around \SI{100}{\micro\second}, the normal retransmission interval is
about \SI{0.2}{ms} (a rate of \SI{5}{kHz}).
With counter values representing about \SI{1}{\micro\second}, 
the retransmit interval would after 900 packets be \SI{2}{ms} (or
\SI{500}{Hz}), which then happens after \SI{0.99}{s}.
After another \SI{99}{s}, the interval would be \SI{20}{ms}.
The maximum RTT value will put a lower rate limit around
\SIrange{1}{10}{Hz} of retransmissions.

\section*{Acknowledgments}
\label{sec:Acknowledgment}
\addcontentsline{toc}{section}{Acknowledgment}

The authors would like to thank the developers of the software tools
GHDL, GTKWave, and Wireshark.

\bibliographystyle{IEEEtran}
\bibliography{IEEEabrv,fnet-article.bib}

\delaytodo[inline]{reference to IEEE 802.3-2018 (ethernet).}
\delaytodo[inline]{(refs for POSIX/BSD sockets and the like seem less urgent to me.}

\end{document}

%% file: fnet-abstract.tex
A common theme of data acquisition systems is the transport of data
from digitising front-end modules to stable storage and online
analysis.  A good choice today is to base this on the ubiquitous,
commercially and cheaply available Ethernet technology.  A firmware
building block to turn already the FPGA of front-end electronics into
a TCP data source and UDP control interface using a data-flow
architecture is presented.  The overall performance targets are to be
able to saturate a 1 Gbps network link with outbound data, while using
few FPGA resources.  The goal is to replace the use of custom data
buses and protocols with ordinary Ethernet equipment.  These
objectives are achieved by being just-enough conforming, such that no
special drivers are needed in the PC equipment interfacing with the
here presented Fakernet system.  An important design choice is to
handle all packet-data internally as 16-bit words, thus reducing the
clock-speed requirements.  An advantageous circumstance is that even
at 1 Gbps speeds, for local network segments, the round-trip times are
usually well below 500 microseconds.  Thus, less than 50 kiB of
unacknowledged data needs to be in-flight, allowing to saturate a
network link without TCP window scaling.  The Fakernet system has so
far been shown to saturate a 100 Mbps link at 11.7 MB/s of TCP output
data, and able to do 32-bit control register accesses at over 450
kword/s.

%% file: figure_topology.tex
\begin{figure}[t]
  \centering
  \includegraphics[width=\linewidth]{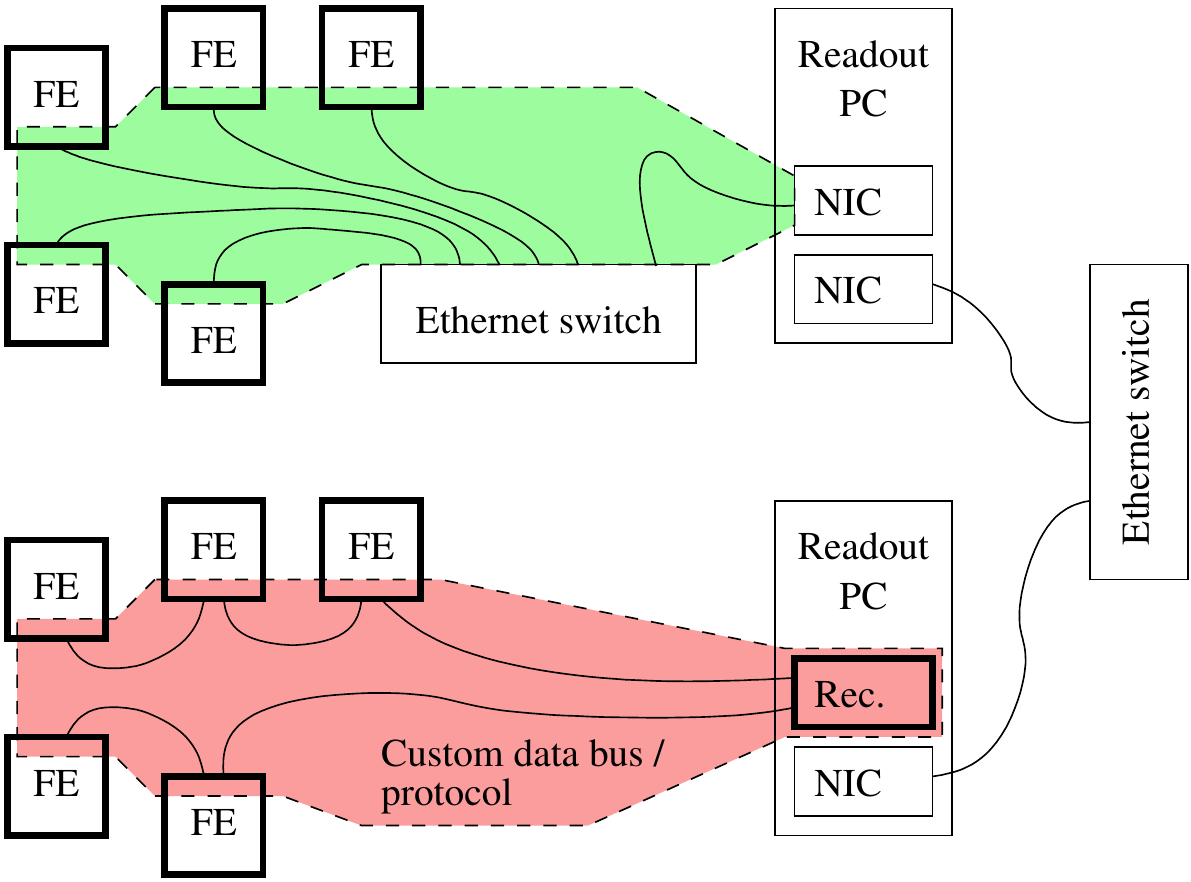}
  \caption{Topology of two readout systems.
    The upper uses Ethernet-enabled front-end (FE) boards together
    with ordinary network equipment for the readout.
    The PC data reception is with a network interface card (NIC).
    The lower uses a custom data transport bus, with associated
    receiver card in the readout PC.
    Both readout PCs have ordinary Ethernet for network access
    to/from the outside.
    Custom hardware is marked with bold outlines.}
  \label{fig:topology}
\end{figure}

%% file: table_custom_buses.tex
\begin{table}[t]
  {
    \centering
    \caption{Some common custom buses used for data acquisition purposes in
      nuclear and high-energy physics experiments.}
    \label{tab:custom_buses}
    \begin{tabular}{l@{}l@{\hskip 0.5\tabcolsep}llll}
      \toprule
      \textbf{Name} & & \textbf{End-point} & \textbf{Bandwidth} &
        \textbf{Links$\times$devices} & \textbf{Ref} \\
       & &  & \textbf{(per link)} & \textbf{per end-point} & \\
       & &  &  (MB/s) &  & \\
      \midrule
      CAEN       & \ldelim\{{2}{*}{}
                   & PCI      & \multirow{2}{*}{85}
                                  & 1$\times$8    & \\[1pt]
      CONET      & & PCIe     &     & 4$\times$8    & \\[1pt]
      \multirow{2}{*}{GSI GTB}
                 & \ldelim\{{2}{*}{}
                   & VME      & \multirow{2}{*}{25}
                                  & 2$\times$31 & \\[1pt]
                 & & PCI      &     & 1$\times$31 & \\[1pt]
      GSI Gosip  & & PCIe     & 200 & 4$\times$255 & \cite{Minami_5750447} \\
      S-LINK$^\mathit{a}$   & & - & 160 & point-to-point & \cite{web:Slink} \\
      S-LINK64$^\mathit{a}$ & & - & 800 & point-to-point &  \\
      \bottomrule
    \end{tabular}\par
    \bigskip
  }
  $^\mathit{a}$The S-LINK specification from CERN describes the interface, but not the
  physical connection, which can vary.
  \delaytodo[inline]{Table/figure should not be on first page.  (IEEE style)}
\end{table}

%% file: table_fpga_tcp_ip_impl.tex
\begin{table*}[t]
  {
    \centering
    \caption{TCP/IP implementations for FPGA end-points.}
    \label{tab:fpga_tcp_ip_impl}
    \begin{tabular}{lllllllll}
      \toprule
      \textbf{Name} & \textbf{Bandwidth} &
         \multicolumn{3}{c}{\textbf{Resource usage}} &
         \textbf{TCP type} & \textbf{Control interface} &
         \textbf{Open} & \textbf{Ref} \\
       & Mbps &
        LUTs & FFs & BRAM &  \\
      \midrule      
      SiTCP & 1000 & 2556 & 3111 & 22 & Bidirectional & UDP protocol & No & \cite{Uchida} \\ %
      
      \textit{Dollas et al.} & 350 & \multicolumn{2}{c}{10007 slices} & 10 & Generic & TCP/UDP raw & Yes & \cite{dollas} \\ %

      \textit{Liu et al.} & 80 (800) & 3134* & 1521 & 4 \si{kiB} & Generic & TCP/UDP raw & ? & \cite{liu} \\ %

      hwNet & 50 & \multicolumn{2}{c}{7265 slices} & 9 & Generic & TCP/UDP raw & ? & \cite{Bergstein}  \\ %

      \textit{Nakanishi et al.} & 95 & 44273 & ? & - & Generic & TCP/UDP raw & ? & \cite{Nakanishi} \\ %

      FEROL & 10000 & $\sim$20000* & ? & & Data source & No & ? & \cite{Bauer_2013} \\ %

      \textit{Sidler et al.} & 10000 & 67938 & 83829 & ? & Generic & TCP/UDP raw & Yes & \cite{Sidler_2015} \\ %
      
      Fakernet & 100 (1000) & 3700 & 2500 & 18 \si{kiB}-- & Data source & UDP `32-bit registers' & Yes & (\cref{table:resource_use})\\ %
      \bottomrule
    \end{tabular}\par
    \bigskip
  }
  \delaytodo[inline]{Another purpose of this table is to show that
    existing implementations are too large.  But lets see what
    we find out there...}
  Cores that only provide UDP and do not implement a TCP sender have
  not been considered, since they would require additional logic to
  provide reliable transport.
  Cores that are TCP offload engines have also not been considered,
  as it defies the purpose of only needing an FPGA.
  Achievable (output) bandwidths are shown.  In two cases, numbers are
  given for perfomance that are claimed to be attainable if using 1
  Gbps interfaces, but have not been tested.
  The resource usage of most designs are given for FPGAs with 4-input LUTs.
  In two cases, the resources are given as slices for Xilinx devices
  that have 2 LUTs per slice.
  In the two cases marked with *, the so-called adaptive LUTs (ALUT)
  of Altera devices are used.

\end{table*}

%% file: figure_matryoshka.tex
\begin{figure}[t]
  \centering
  \includegraphics[width=\linewidth]{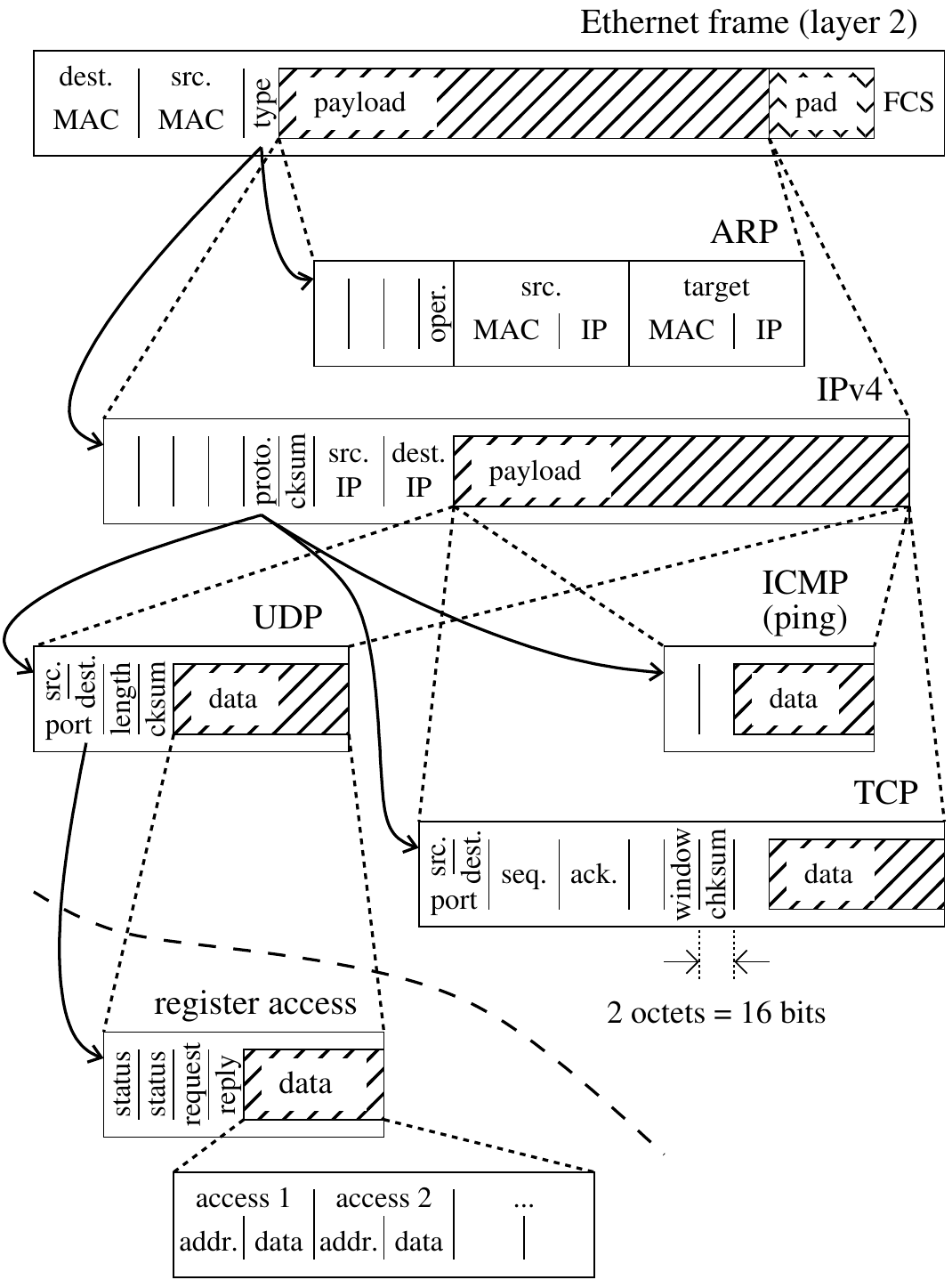}
  \caption{Matroyska doll layout of Ethernet, ARP and IP (TCP, UDP,
    ICMP) packets.
    The more interesting data members are identified with labels.
    Padding is used to reach the minimum ethernet frame length of 64 octets.
    The bottom 'register access' packet is the fakernet-specific
    UDP data format for control operations.
  }
  \label{fig:matryoshka}
\end{figure}

%% file: figure_word_copy.tex
\begin{figure}[t]
  \centering
  \includegraphics[width=\linewidth]{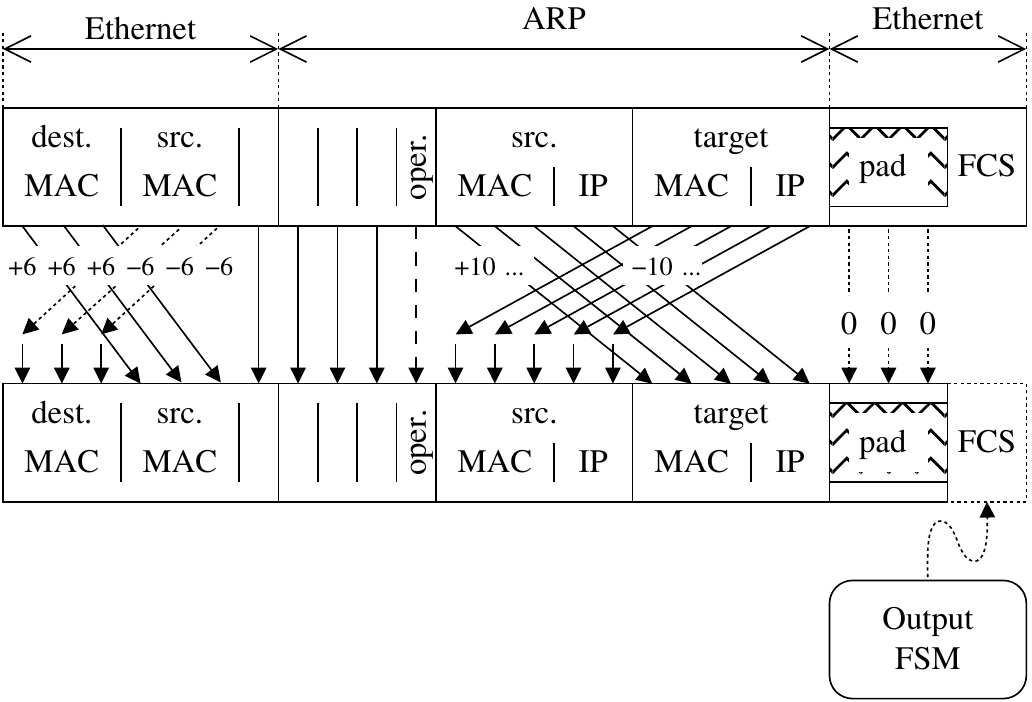}
  \caption{ARP, ICMP and UDP response packets are generated while
    the input packets are parsed, and have the same length.
    The symmetry of source and destination fields (address and port
    numbers) are used,
    by copying or replacing such fields with an offset
    (indicated by the +/- numbers).
    The figure shows the ARP packet case.
  }
  \label{fig:word_copy}
\end{figure}

%% file: figure_overview.tex
\begin{figure*}[t]
  \centering
  \includegraphics[width=\linewidth]{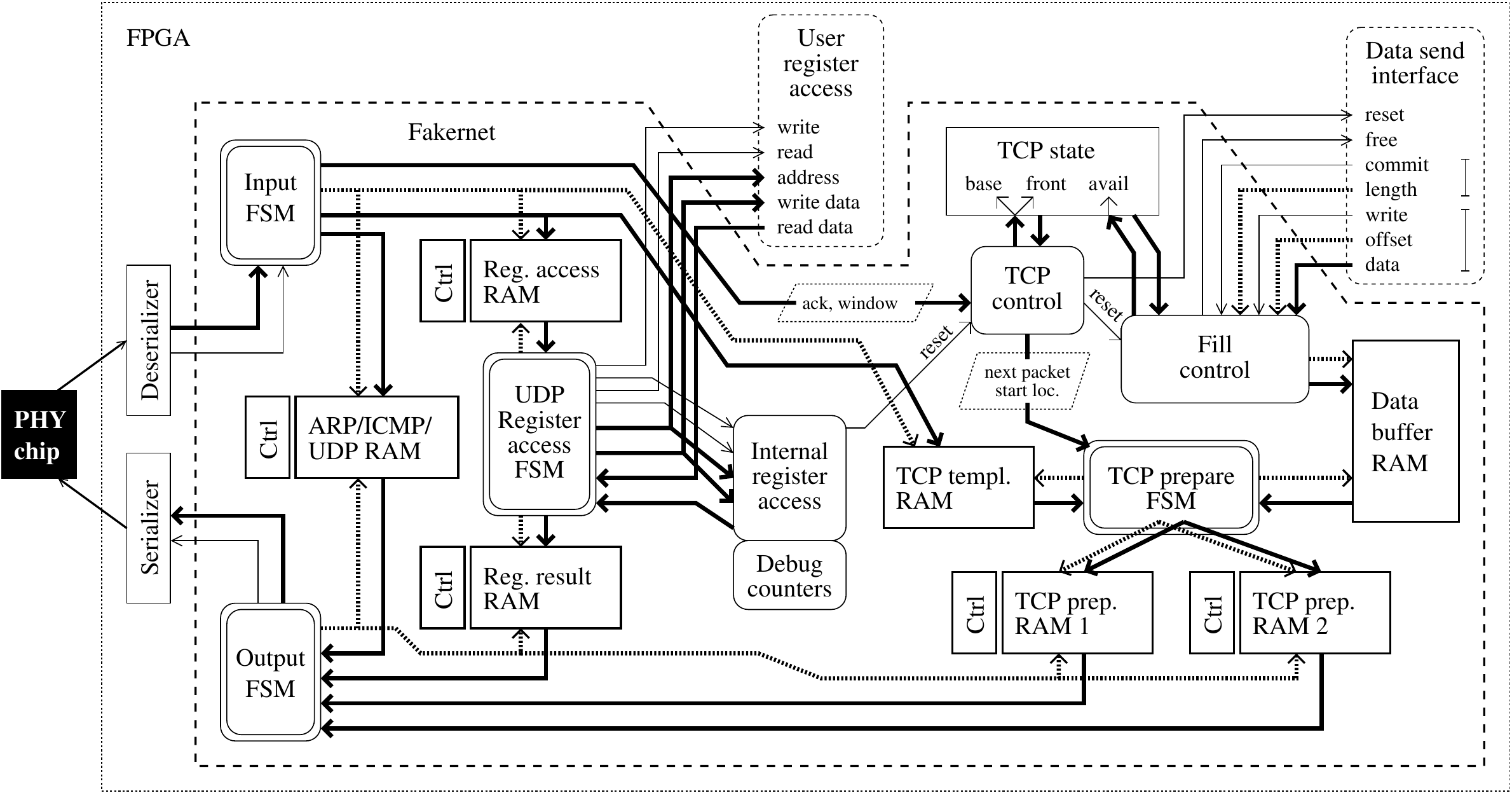}
  \caption{The circuit design consists of a number of finite
    state machines (FSMs) and RAM blocks.
    The FSMs propagate the packets from the input, to the
    RAM blocks and further to the output.
    Together with each RAM block is a control structure which holds
    information on the used size, and whether the block has complete data,
    i.e.\ is ready to be processed by the next stage.
    Each input packet is parsed on-the-fly as it arrives by the input FSM.
    ARP/ICMP response packets are generated directly.
    UDP register access response packet are generated in two steps,
    where the second stage FSM performs the actual register access work
    and fills in the results in the response packet.
    The TCP header of the initial SYN packet is used as template
    for all generated TCP packets.
    After that, the input FSM only give the acknowledged data position
    and window size to the TCP control.
    TCP packets are generated when data to be sent is available,
    or a keep-alive timer has expired.
    Two interfaces connect to other user circuits: the first
    for register access, i.e.\ an address+data interface.
    The second allows the user circuit to provide data to be sent over TCP,
    and is handled by the data buffer fill control.
    The output FSM transmits packets, taken from the different final
    RAM blocks.
    The (de)serialiser at the (input)/output that make up the actual
    interface to the PHY depend on which PHY protocol is used.  }
  \label{fig:overview}
\end{figure*}

%% file: figure_in_fsm.tex
\begin{figure}[t]
  \centering
  \includegraphics[width=0.95\linewidth]{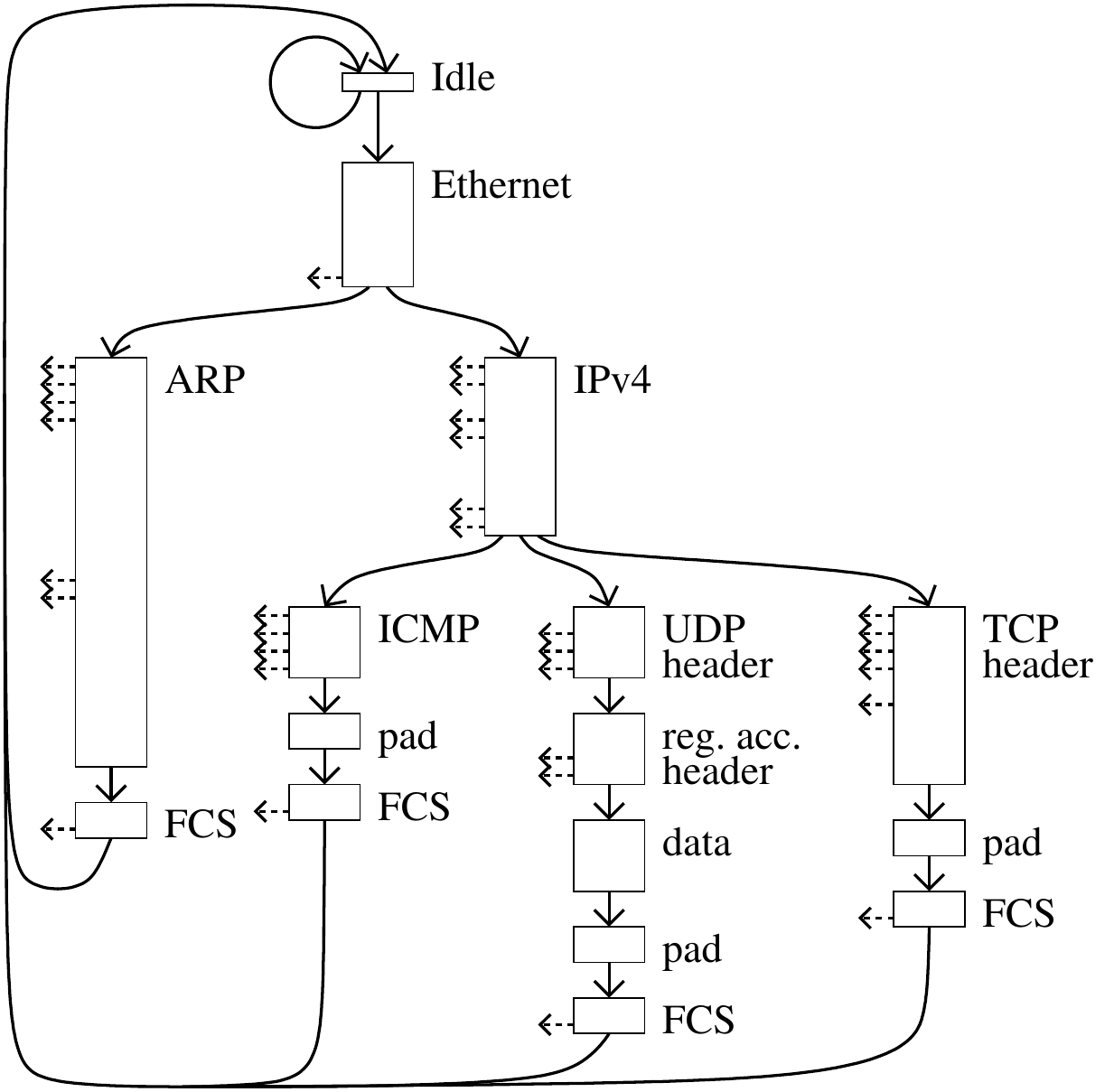}
  \caption{Input packet parsing FSM.
    The handling closely follows the multi-layer packet wrapping of
    \cref{fig:matryoshka}.
    The dashed arrows mark locations where packets can fail
    parsing, transferring the state machine directly
    to the idle state.
  }
  \label{fig:in_fsm}
\end{figure}

%% file: figure_udp_reg_acc.tex
\begin{figure}[t]
  \centering
  \includegraphics[width=0.95\linewidth]{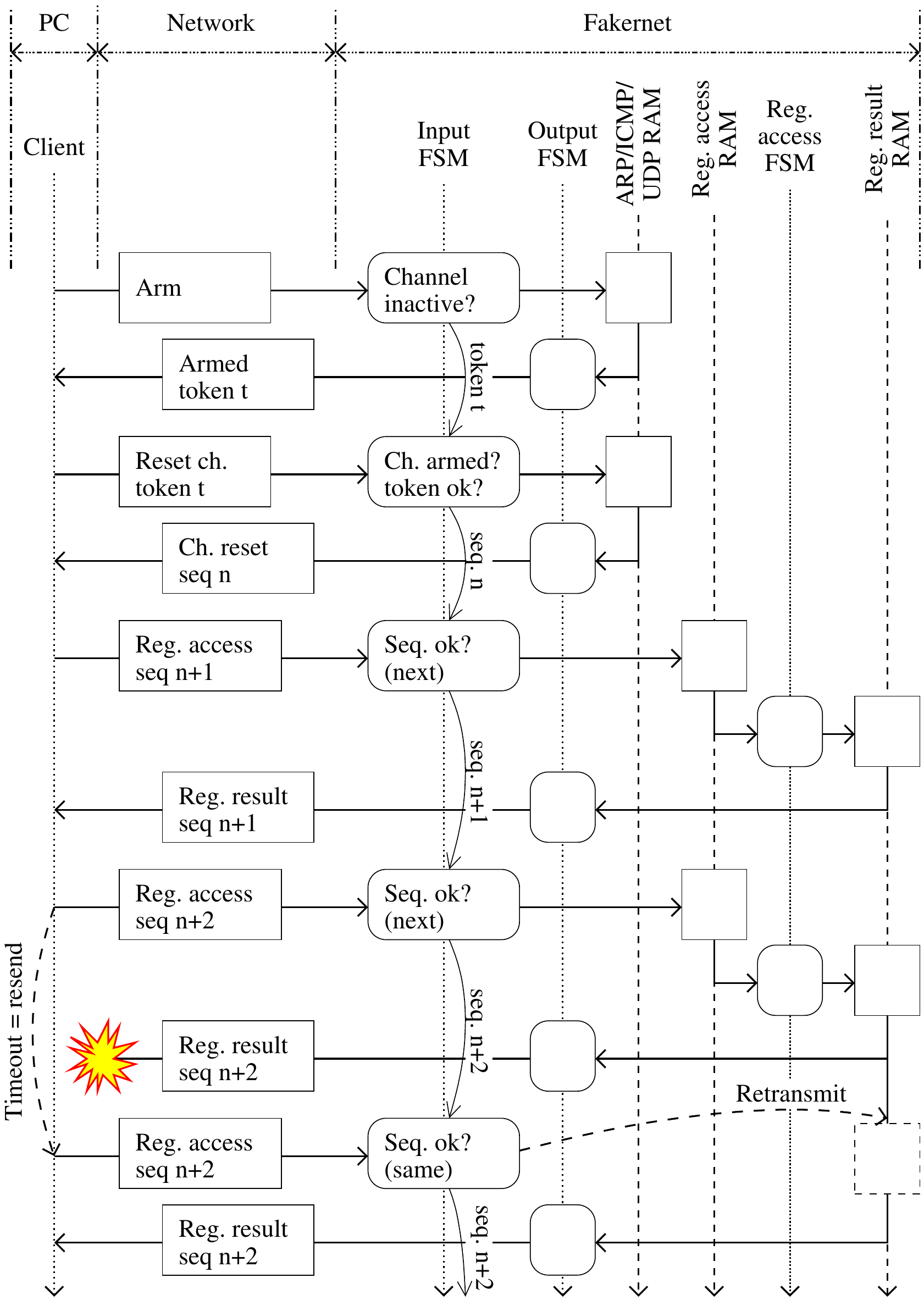}
  \caption{UDP register access protocol example.
    A register access channel is activated by a two-step arm-reset
    sequence.
    The arm response gives a token which must be used
    for the reset.
    The reset response gives the first sequence number for register
    access using the activated channel.
    The IP address and UDP port number of the other end
    are tied after a successful reset.
    Each register access increments the sequence number by one.
    The sequence number thus ensures that each access is only performed
    once.
    The requestor must retry until the response has been received.  }
  \label{fig:udp_reg_acc}
\end{figure}

%% file: figure_data_memory.tex
\begin{figure}[t]
  \centering
  \includegraphics[width=0.75\linewidth]{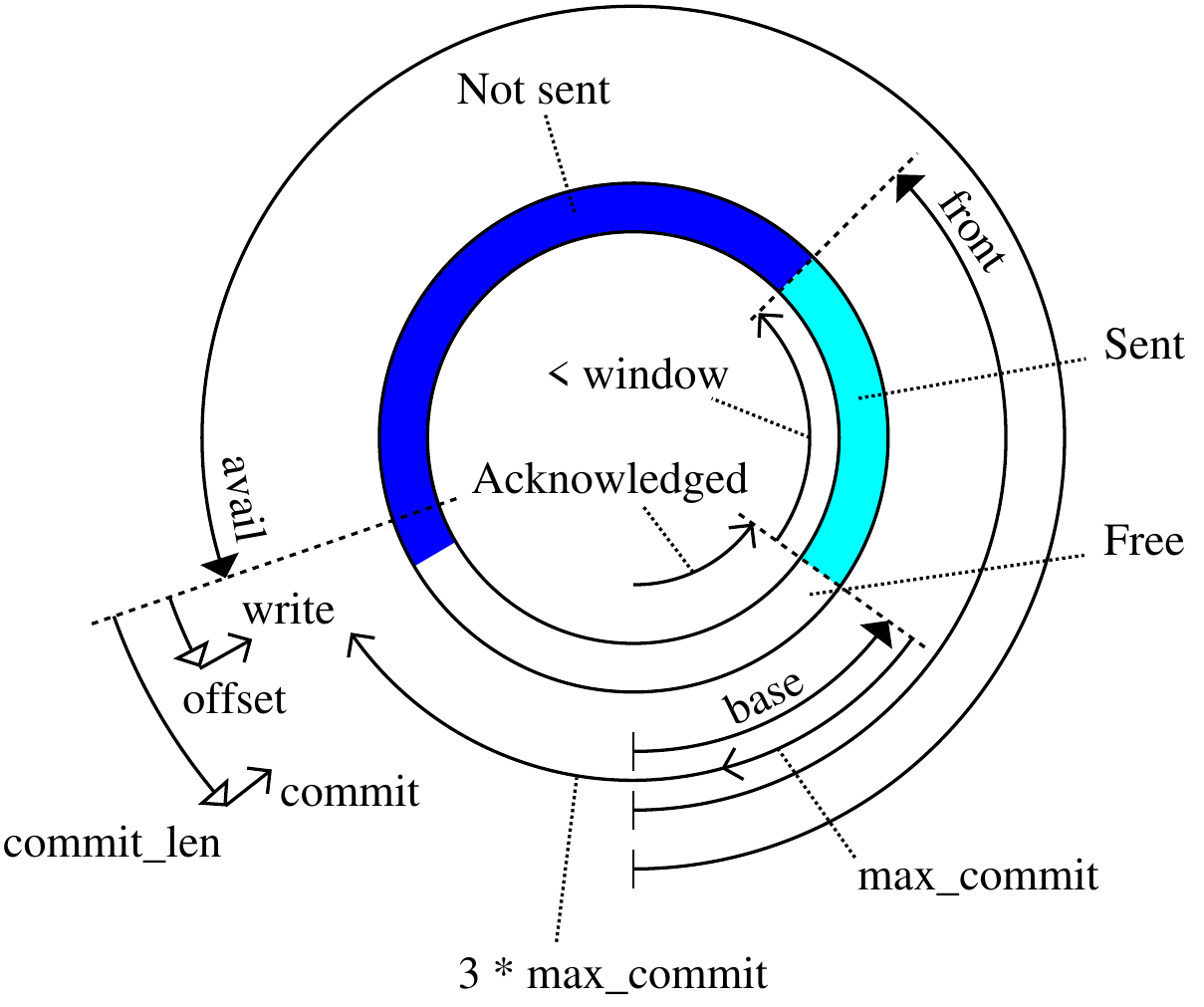}
  \caption{Schematic of the circular data memory.
    The arrows with filled heads give the overall buffer state: \emph{base} is how far
    data has been acknowledged, \emph{front} tells how much data has been
    sent and \emph{avail} how much data is available.
    \emph{front} will never proceed further from \emph{base} than the
    \emph{window} size reported by the receiver.
    The data producer \emph{write}s words at \emph{offset} in the free space,
    and \emph{commit}s data with size \emph{commit\_len}, that then become
    available.
    Free space is not reported to the producer when \emph{avail} has passed the
    point of \emph{-3 max\_commit}, as would be the case after the
    commit in the figure.
    }
  \label{fig:data_memory}
\end{figure}

%% file: table_resource_use.tex
\begin{table*}[t]
  {
    \centering
    \caption{Fakernet circuit resource usage for different FPGA targets,
      and optional components.}
    \label{table:resource_use}
    \begin{tabular}{lcc|ccc|c|c|c|c|ccc|c}
      \toprule
      \multirow{2}{*}{\textbf{Configuration:}} & & &
        \multicolumn{3}{c|}{\textbf{Min}} &
        \multicolumn{1}{c|}{\textbf{Data}} &
        \multicolumn{1}{c|}{\textbf{Debug}} &
        \multicolumn{1}{c|}{\textbf{UDP ch.}} &
        \multicolumn{1}{c|}{\textbf{Buffer}} &
        \multicolumn{3}{c|}{\textbf{Large}} &
        \textbf{Max clock}\\
      \multicolumn{1}{r}{\textbf{}} & \textbf{LUT} & \textbf{RAM} &
        \multicolumn{3}{c|}{\textbf{(2 UDP ch, 4 kiB)}} &
        \multicolumn{1}{c|}{\textbf{gen.}} &
        \multicolumn{1}{c|}{\textbf{reg+cnt}} &
        \multicolumn{1}{c|}{/ch.} &
        \multicolumn{1}{c|}{/addr.bit} &
        \multicolumn{3}{c|}{\textbf{(3 UDP ch, 64 kiB)}} &
        \textbf{frequency}\\
      \midrule
      \textbf{FPGA model} & inputs & \si{kiB} &
        \textbf{LUT} & \textbf{FF} & \textbf{RAM} &
        \textbf{LUT} &
        \textbf{LUT} &
        \textbf{LUT} &
        \textbf{LUT} &
        \textbf{LUT} & \textbf{FF} & \textbf{RAM} &
        {\si{MHz}} \\    
      \midrule
      Xilinx Virtex 4   & 4 & 2 & 2679 & 1650 &  9 &  +429 &  +491 &  +81 &  +11 & 3664 & 2527 & 41 & 131 \\
      \medskip                                                               
      Xilinx Spartan 6  & 6 & 2 & 2423 & 1707 &  9 &  +394 &  +340 & +114 &   +9 & 3259 & 2491 & 41 &  91 \\
      Altera Cyclone V  & A* & 1 & 1157 & 1951 & 20 &  +211 &  +178 &  +58 &   +7 & 1633 & 2821 & 85 & 123 \\
      Altera Max 10     & 4 & 1 & 3568 & 2253 & 20 &  +708 &  +472 & +195 &  +15 & 5085 & 3263 & 84 & 117 \\
      \bottomrule
    \end{tabular}\par
    \bigskip
  }
  In \cref{table:resource_use}, two configurations are described in
  terms of look-up-tables (LUT), signal registers, i.e.\ flip-flops
  (FF), and number of block RAMs used.
  In between, the additional LUT usage for some optional features are given.
  Note that the differences per channel and buffer address bit are
  averages over several increments.
  For both configurations,
  one UDP channel is for idempotent access, i.e.\ usable
  by many clients concurrently.
  The large configuration is not a limit, both a larger buffer as well
  as more UDP access channels can be configured.
  Note that different FPGA models have resources (e.g.\ look-up tables
  (LUT)) with different capabilities, thus the resource consumption
  cannot be meaningfully compared between models.
  The Cyclone V has (adaptive, *) LUTs with variable number of inputs.
  The maximum clock frequency applies at the large configuration,
  but is at most \SIrange{5}{10}{\percent} better for the
  minimum configurations.
\end{table*}

%% file: figure_udp_flood.tex
\begin{figure}[t]
  \centering
  \includegraphics[width=0.9\linewidth]{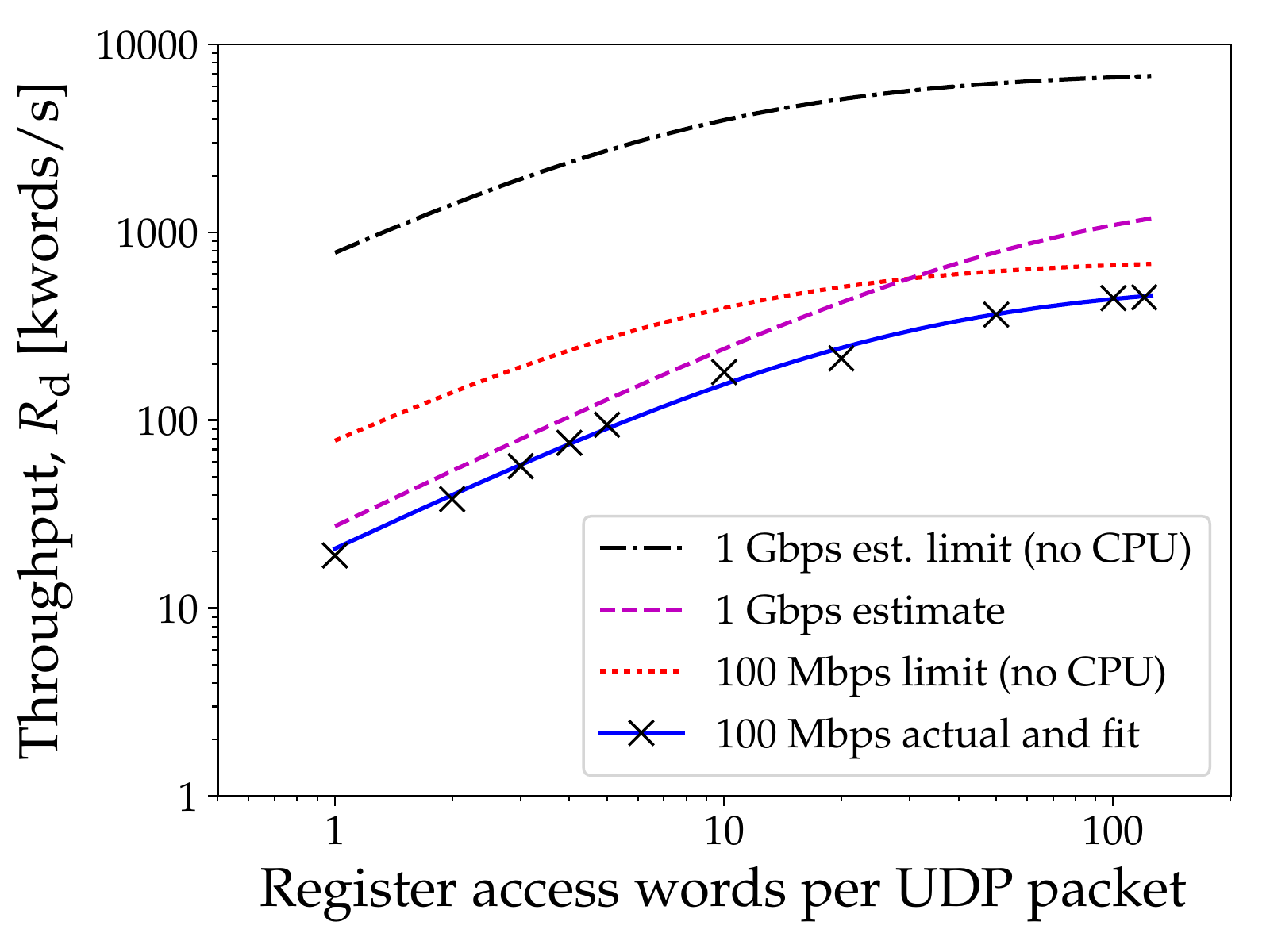}
  \caption{Throughput of the UDP register access protocol,
    as a function of the number of register accesses per packet.
    The dotted line shows the maximally
    attainable throughput, given that new packets are transmitted
    as soon as the previous response has been received (in a
    ping-pong fashion), and assuming immediate response by the FPGA
    code.
    A transferred word is here a 32-bit register access, which uses
    eight octets in both the UDP request and response packets
    (address+data).
    The solid line is a fit including client side (NIC+CPU) overhead of
    \SI{35}{\us/packet} and \SI{0.42}{\us/word}.
    These values also include the effects of one 1 Gbps switch.
    The unavoidable transmission latency at 100 Mbps is
    \SI{11}{\us/packet} and \SI{1.28}{\us/word}.
    The Fakernet access processing adds \SI{0.2}{\us/packet} and
    \SIrange{0.1}{0.2}{\us/word}, depending on how fast the user circuit
    responds.
    The dashed line show extrapolated estimates
    for a 1 Gbps Fakernet interface,
    while the dash-dotted line is the corresponding
    limit without CPU overhead.
  }
  \label{fig:udp_flood}
\end{figure}

%% file: figure_maxpayload.tex
\begin{figure}[t]
  \centering
  \includegraphics[width=0.9\linewidth]{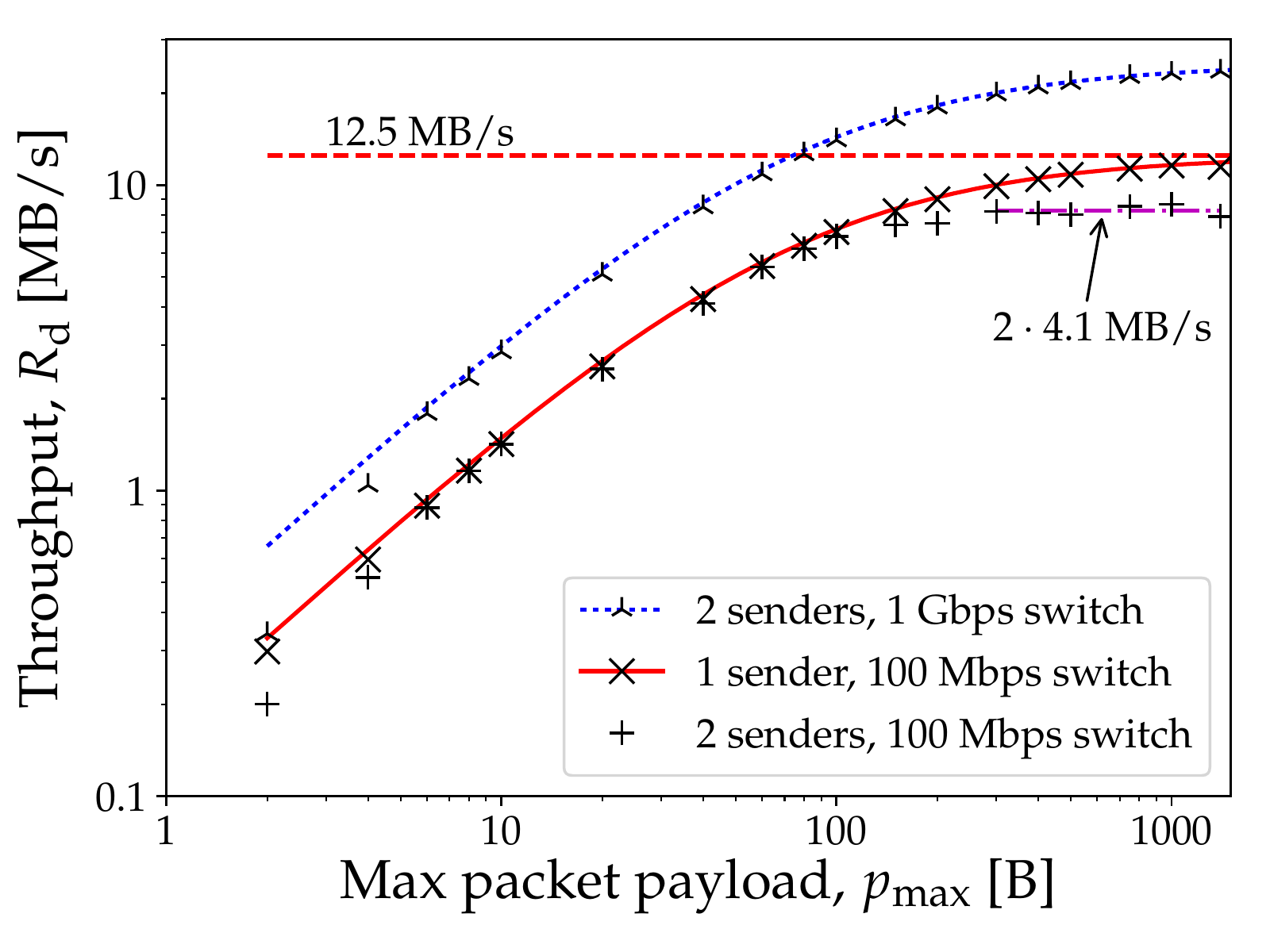}
  \caption{Throughput as function of the maximum TCP data
    payload per packet.
    The dashed line is the maximum possible throughput at \SI{100}{Mbps},
    \SI{12.5}{MB/s} (infinite packet).
    The red solid line shows the expected throughput, which
    is reduced to \SI{11.7}{MB/s}
    due to packet header, preamble, and inter-packet gap overhead,
    with Ethernet frames of the maximum length, 1500 octets.
    Blue dotted is the total throughput for two senders, which
    is twice as large
    when a faster switch that can handle the load is used,
    except for very small payload sizes.
    However, with multiple senders that together try to exceed
    the switch bandwidth (plus signs), the total throughput is reduced.
    (This is expected; such a configuration is not a design target, as it
    violates assumption \ref{nobottleneck} in \cref{sec:simplifying}.)
  }
  \label{fig:maxpayload}
\end{figure}

%% file: figure_maxwindow.tex
\begin{figure}[t]
  \centering
  \includegraphics[width=0.9\linewidth]{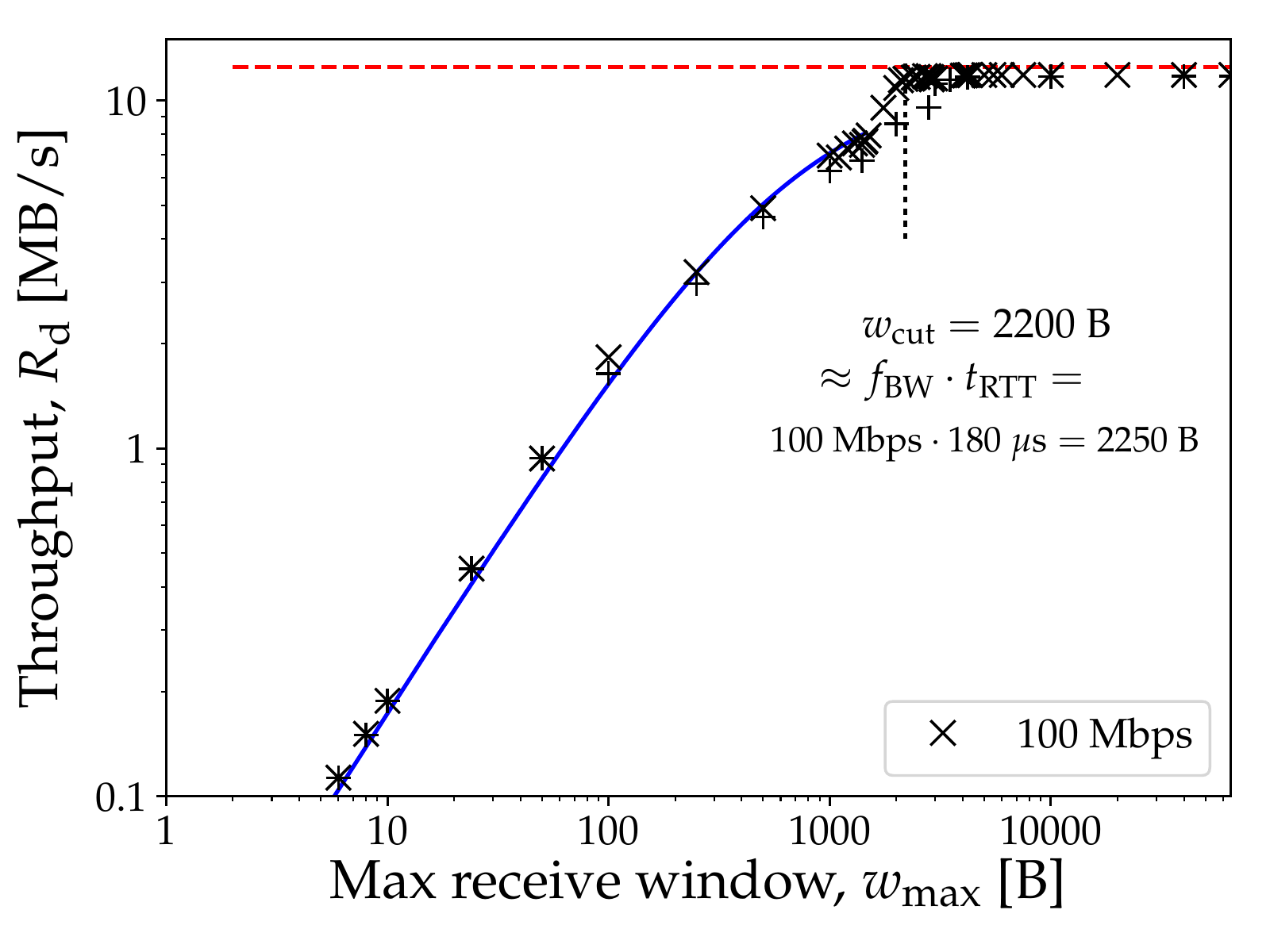}
  \caption{Throughput as a function of an artificially reduced maximum
    receive window.
    When the window size goes below the
    $w_\mathrm{cut}$ threshold, the amount of outstanding non-acknowledged
    data limits the throughput of the connection.
    This is related to the RTT time and link band-width by
    $w_\mathrm{cut} = t_\mathrm{RTT} f_\mathrm{BW}$.
  }
  \label{fig:maxwindow}
\end{figure}

%% file: figure_trickle.tex
\begin{figure}[t]
  \centering
  \includegraphics[width=0.9\linewidth]{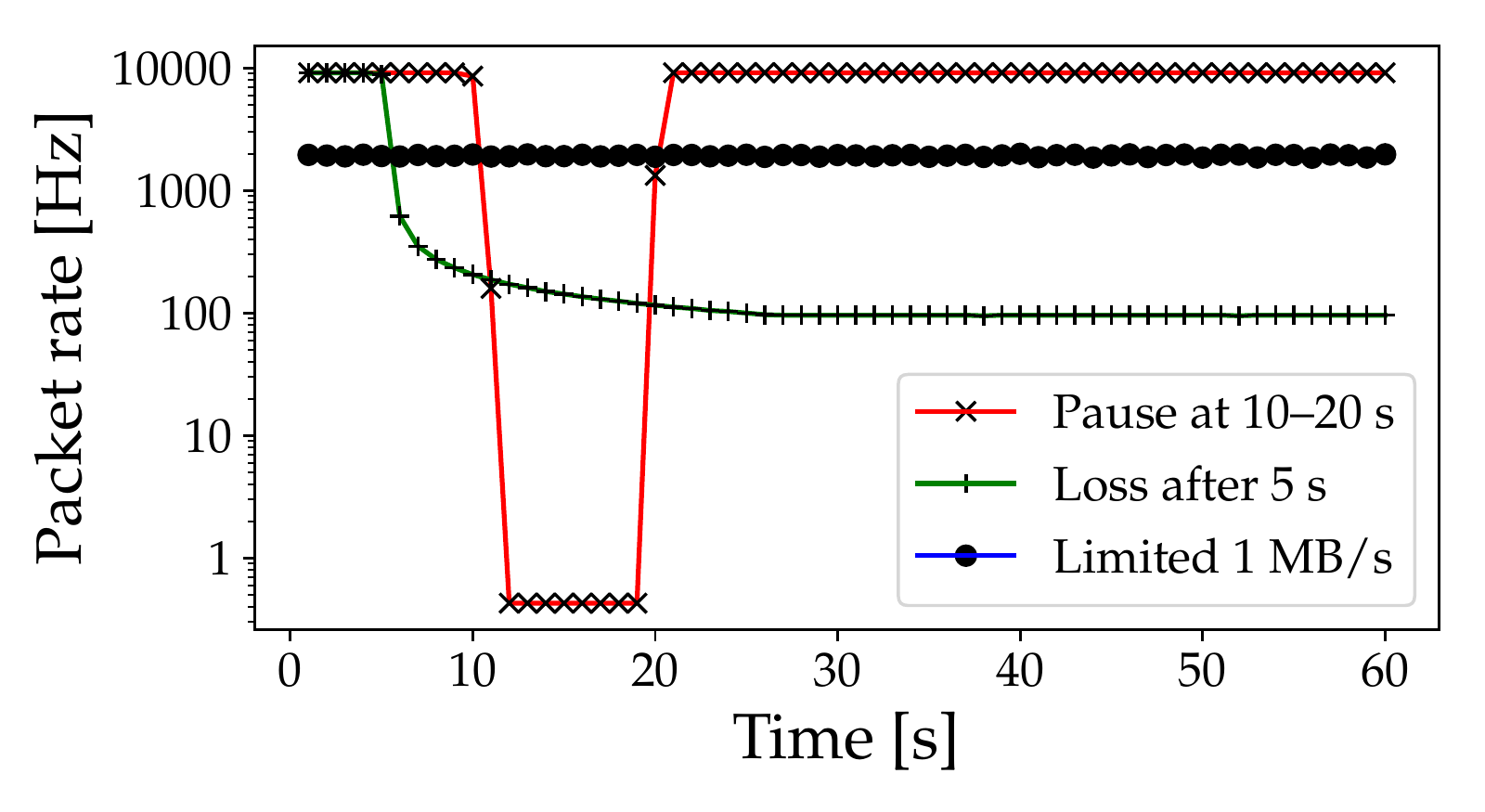}
  \caption{TCP packet rate for three constrained scenarios.
    Crosses with a red line connecting measurements show that the
    reduced (zero) window size is respected for a connection with
    the receiving end pausing reception at \SIrange{10}{20}{s}.
    Only few probe packets are sent.
    Plus signs with a green line shows a connection that is lost
    at \SI{5}{s}, with the following decrease in packet transmissions.
    The circles show a connection where the receiver only accept
    \SI{1}{MB/s}.
  }
  \label{fig:trickle}
\end{figure}